\documentclass[letterpaper, twocolumn,superscriptaddress, showpacs,preprintnumbers,amsmath,amssymb] {revtex4}
\usepackage{graphicx}
\usepackage{dcolumn}
\usepackage{bm}
\usepackage{hyperref}
\hypersetup{
    colorlinks,
    citecolor=red,
    filecolor=blue,
    linkcolor=blue,
    urlcolor=blue
}
\usepackage{amsbsy}

\newcommand{\be}{\begin{equation}}
\newcommand{\ee}{\end{equation}}

\newcommand{\bsig}{\boldsymbol{\sigma}}
\newcommand{\btau}{\boldsymbol{\tau}}

\usepackage{cancel}
\begin{document}

\title{Phenomenology of fully many-body-localized systems}

\author{David A. Huse}
\affiliation{Physics Department, Princeton University, Princeton, NJ 08544, USA}
\author{Rahul Nandkishore}
 \affiliation{Princeton Center for Theoretical Science, Princeton University, Princeton, New Jersey 08544, USA}
\author{Vadim Oganesyan}
\affiliation{Department of Engineering Science and Physics, College of Staten Island, CUNY, Staten Island, NY 10314, USA}
\affiliation{Initiative for the Theoretical Sciences, The Graduate Center, CUNY, New York, NY 10016, USA}
\begin{abstract}
We consider fully many-body localized systems, i.e. isolated quantum systems where all the many-body eigenstates of the Hamiltonian are localized.
We define a sense in which such systems are integrable,
with localized conserved operators.
These localized operators are interacting pseudospins, and the Hamiltonian 
is such that unitary time evolution produces
dephasing but not `flips' of these pseudospins.  As a result, an initial quantum state of a pseudospin can in principle
be recovered via (pseudospin) echo procedures.  We discuss how the exponentially decaying interactions between pseudospins lead to logarithmic-in-time
spreading of entanglement starting from nonentangled initial states.
These systems exhibit multiple different length scales that can be defined from exponential functions of distance;
we suggest that some of these decay lengths diverge at the phase transition out of the fully many-body localized phase while others remain finite.
\end{abstract}
\date{\today}
\maketitle

Isolated quantum many-body systems with short-range interactions and static randomness may be
in a many-body localized phase where they do not thermally equilibrate under their own dynamics.
While this possibility was pointed out long ago by Anderson \cite{pwa}, such localization of highly-excited states
in systems with interactions did not receive a lot of attention
until more recent work \cite{baa, mirlin, agkl, oh, prosen, pal}
brought the subject into focus. Although the original idea of many-body localization came from considering spins in solids \cite{pwa},  more
recent interest in the unitary quantum dynamics of many-body systems fully isolated from their environment
is also due to developments in atomic physics that allow good approximations to such systems to be assembled in the
laboratory, e.g. using systems of cold neutral atoms \cite{bloch} or ions \cite{blatt}. Interest in many-body localization also accrues from the fact that localization can protect types of order that are forbidden in equilibrium \cite{wp, stark, bravyi, z2, bn, pekker, voskprime, chandran, bahri,kjall, qhmbl}, which may have implications for quantum devices and quantum computation.

Isolated systems in the localized phase have strictly
zero thermal conductivity \cite{baa}, so if energy is added to the system locally, it
does not diffuse, even when the system's energy density corresponds to a nonzero
(even infinite \cite{oh}) temperature. Many-body localized energy eigenstates violate the Eigenstate Thermalization Hypothesis
(ETH) \cite{eth1, eth2, eth3} and exhibit only area-law entanglement, unlike the volume-law entanglement of excited eigenstates at nonzero temperature in
thermalizing systems. It is also known \cite{prosen, bard, vosk, iyer, spa1} that for generic initial area-law-entangled states in the many-body localized phase,
the entanglement spreads logarithmically with time, unlike thermalizing systems (where entanglement can spread ballistically \cite{prosen, kh}) and single-particle
localized systems (where the entanglement remains area-law).

In this paper, which is an extended version of Ref. \cite{ho}, we further
explore the phenomenology of {\it fully} many-body localized (FMBL) systems \cite{spa2, bn, bs, ho, bath, arcmp, vos} (systems where {\it all}
the many-body eigenstates of the Hamiltonian display localization).
We argue that there must exist localized pseudospin operators in terms of which the many-body eigenstates within the localized phase are indeed precisely product states with zero entanglement.
The existence of such a construction has recently been proven for a certain class of spin chains \cite{imbrie}.  Writing the Hamiltonian in terms
of these localized pseudospin operators reveals that fully many-body localized Hamiltonians are a type of integrable system, which contain an even
larger number of local conserved quantities than do traditional integrable systems. Additionally, this structure is robust to small but otherwise
arbitrary local perturbations of the Hamiltonian, which only lead to a redefinition of the local constants of motion. We note that when the
Hamiltonian is expressed in terms of these localized pseudospins, it has exponentially decaying long-range interactions which produce dephasing
but {\it do not} produce spin flips. It is these interactions that cause the logarithmic spreading of entanglement observed when the system
is initialized in a nonentangled product state of the bare spins \cite{prosen,bard,vosk,iyer,spa1}. We note also that the effective
interaction between distant pseudospins depends sensitively on the configurations of all intervening pseudospins, and thus changes
from one many-body eigenstate to the next - a form of `chaos' reminiscent of spin glasses \cite{BrayMoore}.

{\it The model}: To be concrete, assume we have a system of $N$ spin-1/2's on some lattice (say, in one, two or three dimensions),
labeled by Pauli operators $\{ \bsig_i\}$. We call these spins `p-bits' (p=physical). Our system has a specific random Hamiltonian $H$ that contains only
short-range interactions and strong enough static random fields on each spin so that,
with probability one in the limit of large $N$, all $2^N$ many-body eigenstates of this $H$ are localized. For an example, see (\cite{pal}).
The discussion should be readily generalizable to local operators with more than two states, to Floquet systems where the Hamiltonian is a periodic function of time, and to systems where the dominant strong randomness is instead the spin-spin interactions rather than random fields.
In the latter, the pseudospins may instead be localized domain wall operators \cite{z2} or spin-exchange operators \cite{vosk}
and the lowest-energy mode may be either a global symmetry mode \cite{z2} or bilocalized between distant sites \cite{vosk}.

We expect that in this fully localized regime, we can define another set of localized Pauli operators $\{\btau_i\}$,
that we refer to as `l-bits' (l=localized) or pseudospins, such that the Hamiltonian when written in terms of these $\btau$ operators takes the form
\begin{equation}
H=\sum_ih_i\tau^z_i + \sum_{i,j}J_{ij}\tau^z_i\tau^z_j + \sum_{n=1}^{\infty} \sum_{i,j,\{k\}}K^{(n)}_{i\{k\} j}\tau^z_i\tau^z_{k_1}...\tau^z_{k_n} \tau^z_j
\end{equation}
with the $K^{(n)}$ terms representing $(n+2)$-l-bit interactions. The sums in (1) are restricted so that each
interaction term is counted only once.  Also we have added a constant to shift the zero of energy (if necessary) so that the trace of $H$ vanishes.
Note that the $\tau^z_i$ all commute with the Hamiltonian and with each other, so the eigenstates of $H$ are simultaneous eigenstates of all the $\tau^z_i$, with zero entanglement of these l-bits.

The intuition underlying the Hamiltonian (1) is that since there is no transport in the localized regime, there should be a set of localized conserved `charges' (the $\{\tau^z_i\}$), which are constants of motion of the system.  E.g., for a system of non-interacting fermions all localized in a disordered potential, the $\{\tau^z_i\}$ would just be the occupation numbers of the localized single-particle orbitals.  Since these l-bits are localized, when written in terms of the p-bits they consist of a sum of terms that are products of p-bit operators on nearby sites, as we discuss below.  These terms have weights that typically fall off exponentially with the distance to the farthest p-bit operator involved in the operator product.  These exponential tails mediate the long-range interactions between l-bits, which thus also fall off exponentially with distance.  The l-bits are thus `dressed' versions of the p-bits, with local `dressing' that makes each $\tau^z_i$ conserved; this dressing also produces the l-bit interactions in $H$.

We will shortly explain how the l-bit operators $\btau$ may be constructed. However, first we discuss how (1) may be used to understand the quantum dynamics in the FMBL regime,
as has been explored in Refs. \cite{prosen, bard, vosk, iyer, spa1}. These works studied real-time dynamics of FMBL systems, starting from
simple initial product states of the bare (p-bit) degrees of freedom.
When written in terms of the l-bits, such p-bit product states have area-law entanglement and thus contain exponentially many eigenstates of $H$.
Importantly, the presence of
interactions between the l-bits
means that such initial states
will dephase, so there will be no local observables that show long-time persistent oscillations.
The
dynamics of the l-bits in the many-body localized phase is in some sense simple: their $z$ components are frozen, while their transverse $xy$ components precess
about the $z$ axes of their Bloch spheres.  However, the precession rate depends on the states of all the other $\tau^z$'s,
due to the interactions between l-bits.
As a result, the $xy$ components of each l-bit become entangled with the $z$ components of all the other l-bits, resulting in 
dephasing and decoherence.
But all the $\tau^z_i$'s are conserved, so there is no `dissipation', and
this dephasing can be reversed by spin echo procedures \cite{echo, Vasseur}.

Next let's consider the spreading of entanglement within the FMBL phase.  As in Refs. \cite{prosen,bard,vosk}, start with an initial state
that is a pure product state of the p-bits.  It follows from our discussion above that such initial states of zero p-bit entanglement generically have extensive diagonal entropy
when expressed in terms of the many-body localized eigenstates of $H$ and the l-bits.
However, this state initially has no entanglement between p-bits,
and is thus a very particular linear combination of the eigenstates of the Hamiltonian.  The eigenstates of the Hamiltonian each have short-range `area-law' entanglement
between the p-bits, while they are product states of the l-bits.  On a microscopic time scale, this initial linear combination of the eigenstates of $H$ will dephase,
producing an area-law entanglement between the p-bits with a magnitude set by the typical entanglement in an eigenstate of $H$, as was seen
in the early time regime in Refs. \cite{bard,vosk}.

After this early-time transient, we can discuss what happens at later times in terms of the l-bits. It is instructive to contrast
with what happens in non-localized, thermalizing many-body systems (see, e.g., \cite{kh}).
In thermalizing systems, the interaction of spins
(p-bits) $A$ and $B$ generates entanglement between spins $A$ and $B$. The subsequent interaction of spins $B$ and $C$ causes $C$ to get entangled not only with $B$,
but also with $A$. As a result, entanglement spreads ballistically, at a speed akin to the Lieb-Robinson speed. However, this ballistic spreading is absent in the
FMBL phase because the interaction between two spins (now l-bits) $B$ and $C$ depends only on their $\tau^z$ values, and the $\tau^z$ value of the spin $B$ is
unaffected by its interaction with the spin $A$ (since $\tau^z$ is a constant of motion). As a result, l-bits can get entangled only through their direct interaction.
 An interaction $J$ has an influence on the phase of a precessing l-bit which becomes significant once $Jt$ is of order one ($\hbar=1$).  Thus, if $J(L)$ is the effective interaction
 at a range $L$, then l-bits separated by a distance $L$ will grow entangled with each other (and with all intervening l-bits) after a time $t \sim 1/J(L)$.
 Since the effective l-bit interactions in the localized phase fall off exponentially with distance, after a time $t$, a given l-bit is entangled with all other l-bits
 within a volume $\sim \log^d{t}$ for a $d$-dimensional system.

More quantitatively, let us define the effective two l-bit interaction $J^{eff}$ in a particular many-body eigenstate as 
\begin{equation}
J^{eff}_{ij} = J_{ij} + \sum_{n,\{k\}}K^{(n)}_{i\{k\} j} \tau^z_{k_1} \tau^z_{k_2} ... \tau^z_{k_n} ~.
\end{equation}
We expect this effective interaction to decay with distance $r$ as $J^{eff}(r) \sim J_0\exp{(-r/\tilde\xi)}$.
This defines an interaction decay length $\tilde\xi$, which will vary over the eigenstates,
as is discussed below.  Note that this effective interaction at distance $r$ is a sum of $\sim 2^r$
interaction terms, so clearly the typical \emph{individual} term in this sum falls off exponentially in $r$ with a shorter
decay length than $\tilde\xi$.  This illustrates that for these systems there are multiple exponential decay
lengths that may behave differently from one another.  The localization length $\xi$ that is expected to diverge at the phase transition
out of the FMBL phase may differ from this interaction length $\tilde\xi$, as is also discussed below.

Let us consider a generic FMBL spin chain with a nonentangled initial pure product state.
If we then consider the long-time growth of the bipartite entanglement entropy between two semi-infinite
half-chains, the distance $x$ that the entanglement spreads in time $t$ is set by $J^{eff}(x) \sim 1/t$, or $x \sim \tilde\xi\log{(J_0t)}$ for the eigenstates with interaction length $\tilde\xi$.  At long time this initial state dephases to produce diagonal entropy per spin $s(\tilde\xi)$ from the eigenstates with $\tilde\xi$.
 The resulting entanglement entropy thus grows as
 $S \sim s(\tilde\xi)\tilde\xi\log{(J_0t)}$, which is dominated at long time not by the eigenstates within the initial state that maximize its diagonal entropy, but by those that maximize the product $s(\tilde\xi)\tilde\xi$.

 The p-bits are composed of local l-bits, so their entanglement will also grow this way at long time.
 This scenario seems consistent with the results reported in Refs. \cite{prosen,bard}. Note that Ref. \cite{vosk} considered a special model at a random-singlet-type critical point within the localized phase \cite{z2}, where the interactions instead fall off with distance as a `stretched exponential', allowing the entanglement to grow as a larger power of $\log t$.

This logarithmic-in-time growth of entanglement can continue without limit in an infinite system, due to the weak long-range interactions
between the l-bits.
Note that the long-time
entanglement entropy per spin will depend on the choice of initial states.  Ref. \cite{bard} chose initial states with the p-bits randomly oriented along
their $z$ axes, which produces a rather small entropy, allowing their DMRG calculation to access fairly long times.  A larger entropy in the same model
at the same energy can be produced by orienting the p-bits initially perpendicular to their $z$ axes \cite{nanduri}.

We now discuss how to obtain (1) starting from a generic p-bit Hamiltonian with strictly short range interactions.
In any system of $N$ p-bits, a construction of $N$ operators $\btau$ which commute with the Hamiltonian can always be made \cite{lych}.  In fact, there are $(2^N)!$ discretely different ways to do it,
since there are that many one-to-one assignments between the $2^N$ many-body eigenstates of $H$ and the $2^N$ simultaneous eigenstates of all of the
$\tau^z_i$'s.  To fully specify such an assignment there are $(2^N -1)$ relative phases between eigenstates that also need to be set.
However, almost all such assignments will fail to produce localized l-bits. Nevertheless, in the localized phase there should be assignments that do produce
localized l-bits.  We now discuss how one may define the `best' such assignment.

For a weakly interacting p-bit Hamiltonian, one can attempt to construct l-bits iteratively, by dressing the p-bit operators so as to ensure commutation
with the Hamiltonian order by order in perturbation theory in the p-bit interactions.  At $n^{th}$ order in perturbation theory the l-bit $\btau_i$ will be a
linear combination of p-bit product operators containing p-bits within a distance $n$ of site $i$. Equivalently, to involve a p-bit a distance $n$
from site $i$ in the definition of $\btau_i$, one must go to order $n$ in perturbation theory. Thus, in the perturbative construction, the l-bits are simply
dressed p-bits, where the weight of the `dressing' falls off exponentially with the distance. Nevertheless, this perturbative construction will ultimately
fail because of degeneracies, which make the definition of l-bits ambiguous \cite{imbrie}.  We thus need a more formal (and non-perturbative) definition of l-bits,
which we now provide.  See also Ref. \cite{LIOM} for an alternative approach to finding l-bit operators.

First, let's look at
one specific location $i$.
Each of the many-body eigenstates of $H$ is specified to be a simultaneous eigenstate of all of the
$\{\tau^z_j\}$'s, with one particular one-to-one assignment, with phases, now assumed.  Of these eigenstates, half have $\tau^z_i=+1$; let's call those states
$\{|\alpha\rangle\}$.  For each of these $2^{(N-1)}$ states $|\alpha\rangle$ we can flip l-bit $i$ to make the state $|\bar\alpha\rangle=\tau^x_i|\alpha\rangle$, which is, by construction, also a many-body eigenstate of $H$ and has $\tau^z_i=-1$, while all the other $\tau^z_j$'s have the same value in $|\alpha\rangle$ and $|\bar\alpha\rangle$.  Thus we can define the l-bit Pauli operators (with the proper commutation relations) at location $i$ as

\begin{eqnarray}
\tau^z_i&=&\sum_\alpha (|\alpha\rangle\langle\alpha|-|\bar\alpha\rangle\langle\bar\alpha|)~,\\
\tau^x_i&=&\sum_\alpha (|\alpha\rangle\langle\bar\alpha|+|\bar\alpha\rangle\langle\alpha|)~,\\
\tau^y_i&=&-i\sum_\alpha (|\alpha\rangle\langle\bar\alpha|-|\bar\alpha\rangle\langle\alpha|)~.
\end{eqnarray}
Note that each $\tau^z_i$ consists of a sum of projectors on to many-body eigenstates of $H$ and thus
commutes with $H$ and with $\tau^z_j$ for all other sites $j$.  To define the l-bit operators at all other locations, just repeat the above.

Next we want to express each such l-bit operator in terms of the bare p-bit operators.  The full set of all linear operators on our $2^N$-dimensional
state space is $4^N$ linearly-independent operators.  One way to list these operators is all $4^N$ composite operators that can made as (outer) products of one
p-bit Pauli operator $\{\sigma^a_i\}$ from each site, where $a=0,x,y$ or $z$, and $0$ denotes the identity operator for that p-bit.  Of these $4^N$ p-bit product operators, only of order $N$ of them are `local' operators that consist of the identity operator at every site except at one or a few sites that are all near each other.  The vast majority are, on the other hand, `global' operators that operate nontrivially and simultaneously on of order $N$ of the p-bits.  For a given Hamiltonian $H$, and a given assignment of all its many-body eigenstates to eigenstates of the l-bits, the l-bit operators as defined above can each be expanded in terms of these p-bit product operators.

Each p-bit product operator has a `range' $\ell$, which can be defined as the distance between the two farthest-apart non-identity local p-bit operators that it contains.  Thus we can define the mean range $\bar\ell_i$ for l-bit $i$, from the weighted (by the norm of the operator) average of the range of all of its
constituent p-bit product operators.  And we can define the average range for a given choice of l-bit operators as the average of the range over all the l-bits.
Of course, other definitions of the average range that are different in their details can be formulated and might be more useful under some circumstances.

We expect that for a generic Hamiltonian in the FMBL regime there do exist definitions of the l-bits that give
a finite average range in the thermodynamic limit.
We want to choose the assignment that minimizes the average range,
and this minimum range will be one measure of the
localization length $\xi$ of the l-bits.  We expect that if we use this optimal assignment, the typical l-bit will consist of an infinite sum of p-bit product operators, but that the terms with long range will
have a total weight that typically falls off exponentially with the range.  Also there will be rare l-bits that have much longer than typical mean range, due to rare `resonances', but these will occur with a probability that falls off exponentially with the range. Similarly, even though the p-bit Hamiltonian contains only short range interactions, the p-bits when expanded in terms of l-bits consist of an infinite sum of l-bit product operators, with long range terms falling off exponentially with range. Thus, the l-bit Hamiltonian (1) will contain interactions between all l-bits, but with the interaction strengths falling off exponentially with the range.

Thus, we have argued that systems in the FMBL regime can be viewed as a type of `integrable'
system, with Hamiltonian (1), which can be used to understand their dynamics. Traditional, translationally-invariant integrable one-dimensional models of $N$ spins have $N$ conserved local densities.  It appears that if you try to make other
conserved quantities as composites (operator products) of these basic conserved densities, these are necessarily nonlocal operators of range $\sim N$.
For a FMBL system, on the other hand, if we consider $n$ l-bits near site $i$, out of products of these l-bits we can make $2^n$ independent
conserved quantities that are all localized near $i$.  In this sense, fully many-body localized systems have many more conservation laws that can affect local observables
than do traditional translationally-invariant integrable systems.
In addition, this structure is robust to arbitrary small local perturbations of the Hamiltonian, which will only make small changes in the definitions of the localized constants of motion.  This again contrasts with traditional integrable systems, which presumably lose their exact integrability under almost all small local perturbations.

We note that our l-bit construction has localization lengths that are present in the {\it Hamiltonian} (1), which set the length scales for the localization of the l-bits and for the exponential decay
of interactions between l-bits. However, intuition informed by single-particle localization suggests that the typical localization length should vary with energy, and in
particular that the localization length should be longer in the middle of the spectrum
where the many-body density of states is maximal,
and shorter near the edges of the spectrum where the density of states is (exponentially) lower.
Since all eigenstates are eigenstates of the same l-bit Hamiltonian (1), we suggest that the apparent localization lengths in the Hamiltonian are set by the eigenstates that have the longest localization lengths, and a type of `screening' can reduce the localization lengths for other eigenstates.

Next, we discuss how different eigenstates can have different localization lengths $\xi$.
Consider the process of perturbatively `dressing' the p-bits to make the l-bits, and thus diagonalizing the Hamiltonian.  The size of the perturbative effects,
and thus the strength of the dressing and of the long-range l-bit interactions generated by this dressing depends on the ratios of matrix elements to energy denominators that are encountered
in the perturbation series.  The distribution of the magnitude of these ratios will vary between many-body eigenstates, and this allows
different eigenstates to have different localization lengths, $\xi$.  At energies near the center of the many-body spectrum where the density of states
is very large, there can be many other states encountered perturbatively which are close in energy, thus producing smaller energy denominators and larger
perturbative effects.  Near the many-body ground state energy, on the other hand, the density of states is exponentially smaller, and as a result there
should be fewer small energy denominators encountered in the perturbation series.  We expect that $\xi$ will depend on more than just the energy,
since even at a given total energy, one can ask for the eigenstates whose detailed configuration is such that the dressing is
either minimized or maximized.
The eigenstates with the longest localization lengths will presumably be those where the dressing is maximized as much as is possible.  A FMBL system is one where these eigenstates remains localized in spite of the resulting strong quantum fluctuations.

The effective interaction between two distant l-bits is a sum (2) with a number of terms that is exponential in the distance.  Each term has a magnitude that
is set by the Hamiltonian and is the same for each eigenstate.  Only the signs of the terms change between eigenstates.  Thus what must happen is
that the degree of cancelation between these terms must vary among eigenstates, allowing them to have different interaction lengths $\tilde\xi$.
This is a type of `screening' of the l-bit interactions that may allow this interaction length $\tilde\xi$ to be in some cases much less than the
localization length $\xi$ that sets the size of the l-bit operators.  It seems possible that $\tilde\xi$ may even remain finite at the phase
transition out of the FMBL phase where $\xi$ diverges.
It will be of interest to understand this screening process in more detail.

For some less strongly disordered models, we expect that there is a mobility edge within the many-body spectrum of $H$ \cite{baa}.
In the non-localized portion of the spectrum
the many-body eigenstates are expected to obey the ETH \cite{eth1,eth2,eth3}.
For such systems, we expect that all of the above
possible definitions of the l-bits will produce average ranges of order the system size.
Such Hamiltonians are not integrable in any useful sense \cite{lych}, even though we can still formulate a definition of an extensive set of (global) conserved quantities out of the projections on to the eigenstates.
Whether our l-bit construction can be modified to usefully describe the localized regime of such systems remains an open question. It seems possible that some definition of l-bits could exist,
perhaps involving operators somehow projected onto the localized subspace.
But there are difficulties with this idea: one is that typical MBL eigenstates in such systems do have rare regions where the local energy density approaches arbitrarily close to the mobility edge (a new type of `Griffiths singularity'). Similar issues have also been recently discussed in Ref.\onlinecite{vos}. We leave this challenge for future work.

We are grateful to A. Pal, T. Spencer, J. Imbrie, E. Altman, I. Cirac, J. Moore, F. Essler, G. Refael, S. Sondhi, M. Mueller, R. Vosk, D. Abanin, A. Chandran, S. Gopalakishnan, B. Altshuler, I. Aleiner and D. Basko for enlightening discussions.  This work was supported in part by NSF under DMR-0819860 (DAH) and DMR-0955714 (VO), and by funds from the DARPA optical lattice emulator program (DAH).


\begin{thebibliography}{99}
\bibitem{pwa}
P. W. Anderson, Phys. Rev. {\bf 109}, 1492 (1958).

\bibitem{baa}
D.~M. Basko, I.~L. Aleiner and B.~L. Altshuler, Annals of Physics
{\bf 321}, 1126 (2006); cond-mat/0602510 .

\bibitem[AKGL (1997)]{agkl}
B. L. Altshuler, Y. Gefen, A. Kamenev and L. S. Levitov, Phys. Rev. Lett. 78, 2803 (1997).

\bibitem[Mirlin (2006)]{mirlin}
I. V. Gornyi, A. D. Mirlin and D. G. Polyakov, Phys. Rev. Lett. 95, 206603 (2005).

\bibitem{oh}
V. Oganesyan and D. A. Huse, Phys. Rev. B {\bf 75}, 155111 (2007).

\bibitem{prosen}
M. Znidaric, T. Prosen and P. Prelovsek, Phys. Rev. B {\bf 77}, 064426 (2008).

\bibitem{pal}
A. Pal and D. A. Huse, Phys. Rev. B {\bf 82}, 174411 (2010).

\bibitem{bloch}
I. Bloch, J. Dalibard and W. Zwerger, Rev. Mod. Phys. {\bf 80}, 885 (2008).

\bibitem{blatt}
R. Blatt and C. F. Roos, Nature Physics {\bf 8}, 277 (2012).

\bibitem{wp}
J. R. Wootton and J. K. Pachos, Phys. Rev. Lett. {\bf 107}, 030503 (2011).

\bibitem{bravyi}
S. Bravyi and R. Koenig, Comm. Math. Phys., 316 (3), 641-692 (2012)

\bibitem{stark}
C. Stark, L. Pollet, A. Imamoglu and R. Renner, Phys. Rev. Lett. {\bf 107}, 030504 (2011).

\bibitem{z2}
D. A. Huse, R. Nandkishore, V. Oganesyan, A. Pal and S. L. Sondhi, Phys. Rev. B {\bf 88}, 014206 (2013).

\bibitem{bn}
B. Bauer and C. Nayak, J. Stat. Mech., P09005 (2013).

\bibitem{pekker}
D. Pekker, G. Refael, E. Altman, E. Demler and V. Oganesyan, Phys. Rev. X {\bf 4}, 011052 (2014).

\bibitem{voskprime}
R. Vosk and E. Altman, Phys. Rev. Lett. {\bf 112}, 217204 (2014).

\bibitem{bahri}
Y. Bahri, R. Vosk, E. Altman and A. Vishwanath, arXiv:1307.4092.

\bibitem{chandran}
A. Chandran, V. Khemani, C. R. Laumann and S. L. Sondhi, Phys. Rev. B. {\bf 89}, 144201 (2014).

\bibitem{kjall}
J. A. Kj\"all, J. H. Bardarson and F. Pollman, arXiv:1403.1568.

\bibitem{qhmbl}
R. Nandkishore and A. C. Potter, arXiv:1406.0847.

\bibitem{eth1}
J. M. Deutsch, Phys. Rev. A {\bf 43}, 2046 (1991).

\bibitem{eth2}
M. Srednicki, Phys. Rev. E {\bf 50}, 888 (1994).

\bibitem{eth3}
M. Rigol, V. Dunjko and M. Olshanii, Nature {\bf 452}, 854-858 (2008).

\bibitem{bard}
J. H. Bardarson, F. Pollmann and J. E. Moore, Phys. Rev. Lett. {\bf 109}, 017202 (2012).

\bibitem{vosk}
R. Vosk and E. Altman, Phys. Rev. Lett., {\bf 110}, 067204 (2013).

\bibitem{spa1}
M. Serbyn, Z. Papic and D. A. Abanin, Phys. Rev. Lett. {\bf 110}, 260601 (2013).

\bibitem{iyer}
S. Iyer, V. Oganesyan, G. Refael, D. A. Huse, Phys. Rev. B {\bf 87}, 134202 (2013).

\bibitem{kh}
H. Kim and D. A. Huse, Phys. Rev. Lett. {\bf 111}, 127205 (2013).

\bibitem{ho}
D. A. Huse and V. Oganesyan, arXiv:1305.4915.

\bibitem{spa2}
M. Serbyn, Z. Papic and D. A. Abanin,
Phys. Rev. Lett. {\bf 111}, 127201 (2013).

\bibitem{vos}
V. Ros, M. M\"uller and A. Scardicchio, arXiv:1406.2175.

\bibitem{bs}
B. Swingle, arXiv:1307.0507.

\bibitem{bath}
R. Nandkishore, S. Gopalakrishnan and D. A. Huse, arXiv:1402.5971.

\bibitem{arcmp}
R. Nandkishore and D. A. Huse, arXiv:1404.0686.

\bibitem{imbrie}
J. Z. Imbrie, arXiv:1403.7837.

\bibitem{BrayMoore}
A. J. Bray and M. A. Moore, Phys. Rev. Lett. {\bf 58}, 57 (1987).

\bibitem{echo}
M. Serbyn, M. Knap, S. Gopalakrishnan, Z. Papic, N. Y. Yao, C. R. Laumann, D. A. Abanin, M. D. Lukin and E. A. Demler, arXiv:1403.0693.

\bibitem{Vasseur}
R. Vasseur, S. A. Parameswaran and J. E. Moore, arXiv:1407.4476.

\bibitem{nanduri}
A. Nanduri, H. Kim and D. A. Huse, Phys. Rev. B [in press], arXiv:1404.5216.

\bibitem{lych}
O. Lychkovskiy, Phys. Rev. A {\bf 87}, 022112 (2013).


\bibitem{LIOM}
A. Chandran, I. H. Kim, G. Vidal and D. A. Abanin, arXiv:1407.8480.

\end{thebibliography}
\end{document}